\newcommand{\TextPageLimit}{13}
\newcommand{\PageLimit}{18}

\newif \ifhyperlinks    \hyperlinkstrue
\newif \ifDraft         \Draftfalse
\newif\ifFinal \Finaltrue

\documentclass[letterpaper,10pt,twocolumn]{article}
\usepackage[pdftex]{graphicx}
\usepackage{usenix}

\usepackage{fixmath,times,multirow,array}
\usepackage[square,sort&compress,numbers]{natbib}
\renewcommand{\cite}[1]{\citep{#1}}

\usepackage{paralist}           % for compact lists
\usepackage{authblk}

\usepackage{verbatim}           % for comment environment (eats up contents)

\usepackage{xspace}             % to fix spacing in macros producing words. Eg:
\usepackage{balance}            % to balance columns on the last
\usepackage[pdftex]{graphicx}
\setkeys{Gin}{keepaspectratio=true,clip=true,draft=false,width=\linewidth}
\graphicspath{{./imgs/}}
\usepackage{color,soul}
\ifDraft
  \newcommand{\Comment}[1]{\textbf{\textsl{#1}}\xspace}
  \newenvironment{LongComment}[1] % multi-paragraph comment, argument is owner
    {\begingroup\par\noindent\slshape \textbf{Begin Comment[#1]}\par}
    {\par\noindent\textbf{End Comment}\endgroup\par}
  \newcommand{\FIXME}[1]{\textcolor{red}{\textbf{\textsl{FIXME: #1}}}\xspace}
  \newcommand{\TODO}[1]{\textcolor{red}{\textbf{\textsl{TODO: #1}}}\xspace}
  
\else
  \newcommand{\Comment}[1]{\relax}
  
  \newcommand{\FIXME}[1]{\relax}
  \newcommand{\TODO}[1]{\relax}
  
\fi
\newcommand{\CComment}[2]{\textcolor{#1}{\Comment{#2}}}

\renewcommand{\paragraph}[1]{\vspace{1ex}\par\noindent\textbf{#1}\xspace}

\newcommand{\calC}{{\cal C}}
\newcommand{\Cmax}{\calC_0}

\newcommand{\stats}[2]{$\mathbold{n=#1,\calC=#2\,\mbox{b}}$}
\newcommand{\statsm}[3]{$\mathbold{n=#1,\calC=#2\,\mbox{b},\Cmax=#3\,\mbox{b}}$}

\usepackage{xcolor}
\definecolor{linkcolor}{rgb}{0.65,0,0}
\definecolor{citecolor}{rgb}{0,0.4,0}
\definecolor{urlcolor}{rgb}{0,0,0.65}
\usepackage[pdftex, colorlinks=true, linkcolor=linkcolor, urlcolor=urlcolor, citecolor=citecolor]{hyperref}
\hypersetup{pdfpagemode=UseNone,pdfpagelayout=OneColumn,pdfstartview=FitH}
\ifhyperlinks\else   
  \hypersetup{nolinks=true}
\fi

\begin{document}
  \sloppy

  %% Capitalisation of cross references
  \renewcommand{\sectionautorefname}{Section}
  \renewcommand{\subsectionautorefname}{Section}
  \renewcommand{\subsubsectionautorefname}{Section}
  \renewcommand{\appendixautorefname}{Appendix}
  \renewcommand{\Hfootnoteautorefname}{Footnote}
  %% Commands for index
  \newcommand{\Htextbf}[1]{\textbf{\hyperpage{#1}}}

\title{Your Processor Leaks Information -- and There's Nothing You Can Do About It}

\author[1,2]{Qian Ge}
\author[1,3]{Yuval Yarom}
\author[1,2]{Frank Li}
\author[1,2]{Gernot Heiser}
\affil[1]{Data61, CSIRO}
\affil[2]{UNSW, Australia}
\affil[3]{School of Computer Science, The University of Adelaide}
\affil[ ]{\textit {\{qian.ge,yuval.yarom,frank.li,gernot.heiser\}@data61.csiro.au}}

  \maketitle

  \subsection*{Abstract}
 Timing channels are information flows, encoded in the relative timing
of events, that bypass the system's protection mechanisms. Any
microarchitectural state that depends on execution history and
affects the rate of progress of later executions potentially
establishes a timing channel, unless explicit steps are taken to
close it. Such state includes CPU caches, TLBs, branch predictors
and prefetchers; removing the channels requires that the OS can
partition such state or flush it on a switch of security
domains. We measure the capacities of channels based on these
microarchitectural features on several generations of processors
across the two mainstream ISAs, x86 and ARM, and investigate the
effectiveness of the flushing mechanisms provided by the respective
ISA. We find that in all processors we studied, at least one
significant channel remains. This implies that closing all timing
channels seems impossible on contemporary mainstream processors.

\ifFinal
  \pagestyle{empty}
\else
  \pagestyle{plain}
  \thispagestyle{plain}
\fi

\section{Introduction}\label{s:intro}

Computer hardware is increasingly being shared between
multiple, potentially untrusted, programs.
Examples of such sharing range from cloud services, where
a single computer may share workloads of multiple clients,
to mobile phones that run apps authored by
different developers, to web browsers executing Javascript code
originating from different sites.
To protect confidential or private information that some
of these programs may access,
the system imposes a \emph{security policy} that is designed to prevent
information flow between different security domains (eg.\ VMs, apps or
web pages).

One long-established threat to the security of shared systems are
\emph{covert channels}~\cite{Lampson_73}, which
allow colluding programs
to bypass the security policy, by transferring information 
over media that are not controlled by
the system.
A typical scenario includes two programs:
a \emph{Trojan} program as the sender, which has access to
sensitive information but is confined~\cite{Lipner_75} by the security
policy (i.e.\ prevented from sending information to arbitrary destinations),
and a \emph{spy} program as the receiver, which does not have access to the
sensitive information but can communicate with less
restrictions.

Traditionally, covert channels were considered mainly in the context
of military-grade multi-level  secure systems~\citep{Landwehr_81, DoD_85:orange}.
However, with the spread of cloud platforms and the proliferation of
untrusted mobile code, including mobile-phone apps and third-party
code running in web browsers, covert channels are becoming a
mainstream security problem.
Colluding mobile apps present a concrete example of the risk: Consider an app
which operates on sensitive private data and is denied network access
to prevent it from leaking the data. Such an app (Trojan) could use a covert
channel to leak to a separate app (spy), which has unrestricted
network access but no direct access to sensitive data. Similarly, a
cloud-based web service might contain a Trojan which leaks secret data
to a co-located spy VM, circumventing the encryption of client-server traffic.

Covert channels are usually classified as either
\emph{storage} or \emph{timing} channels~\cite{Schaefer_GLS_77}.
Storage channels represent information as some system state affected by the
sender that can be sensed by the receiver, frequently exploiting operating
system (OS) metadata as the storage medium. 
Past research has demonstrated that storage channels
can be eliminated completely~\cite{Murray_MBGBSLGK_13}.

Here we focus on the open problem of timing channels, which exploit timing variations
for communication. They are harder to deal with, partially because of
the breadth of exploitable mechanisms. For example,
secret-dependent code paths or data access patterns can lead to timing
variations that can be exploited locally~\cite{Osvik_ST_06,Yarom_Falkner_14}
or even remotely~\cite{Bernstein_05},
and the usual defence is to strive for deterministic execution time
via constant-time algorithms~\cite{Bernstein_LS_12}.

Such timing channels are arguably a problem of user-level software,
although OS mechanisms may be employed to ensure deterministic timing
of externally-visible effects~\cite{Askarov_ZM_10,Cock_GMH_14,Braun_JB_15}. 
However, there are other classes
of timing channels that clearly fall into the responsibility of the
OS.\footnote{We use the term ``operating system'' in a generalised
	sense, referring to the most privileged software level that has full
	control over the hardware. In a cloud scenario this would refer to
	the hypervisor, and the term ``process'' would represent a virtual
	machine.} This specifically includes timing channels resulting from
microarchitectural state, such as caches, that is not explicitly
exposed to user-level software. Such channels, from now on simply
referred to as \emph{microarchitectural channels}, are the topic of
this paper.

Specifically, \emph{we examine the degree to which it is possible to prevent
	microarchitectural timing channels on contemporary hardware.} Timing channels
between concurrent executions on a single core (i.e.\ simultaneous
multithreading, SMT) are well-documented and
understood~\cite{Aciicmez_Seifert_07}, are high-bandwidth, and are
probably impossible to close, so we assume a security-conscious OS
that disables SMT. We further restrict our examination to intra-core timing
channels, i.e.\ between time-multiplexed users of a single core.

One might assume that elimination of microarchitectural channels in
this restricted scenario would be trivially (although expensively)
achieved by flushing all microarchitectural state on each switch of
security context. However, as we will demonstrate, this assumption is
wrong on contemporary mainstream hardware, and significant channels
remain despite the best efforts of the OS. In short, \textbf{our
	findings mean that contemporary processors are inherently insecure,
	as they cannot be prevented from leaking information}.

We make the following contributions:
\begin{compactitem}
	%\item We identify a limited scenario for investigating microarchitectural-timing-channel elimination. (\autoref{s:model}.)
	\item We implement covert channels attacking all known caching
	microarchitectural features (\autoref{s:channels}), and
	identify and implement all known mitigation techniques for those
	channels that do not depend on undocumented/unreliable information
	(\autoref{s:mitigations}).
	\item We measure the capacity of those channels with and without
	mitigations on multiple generations of recent implementations of the
	two mainstream architectures, x86 and ARM (\autoref{s:results-sum}).
	\item We demonstrate that on each investigated processor, there are
	significant residual channels that cannot be closed with known
	techniques (\autoref{s:results-sum}).
	% Evtyushkin and others have demonstrated the BTB.
	% \item We are the first to demonstrate using TLB (\autoref{s:tlb-channel}) and BTB (\autoref{s:branch-target-buffer-channel}) as covert channels.  
	\item We find that an apparent I-cache remnant channel
	on x86 as well as ARM
	seems to result from instruction prefetching (\autoref{s:discovery1}). 
	\item We show that the branch target buffer on x86
	provides timing channels that cannot be mitigated
	(\autoref{s:btb-timing-channel-x86}).
	\item We show the branch history buffer provides a channel that cannot
	be closed on x86 platforms and on more recent ARM platforms (\autoref{s:d-branch-channel-result}).
	\item We demonstrate that flushing the TLB is insufficient for
	mitigating the TLB channel on x86 (\autoref{s:discovery4})
	\item We argue the need for a new hardware-software contract,
	extending the ISA with mechanisms required to enforce
	confidentiality on time-shared cores (\autoref{s:discussion}).
\end{compactitem}

Our results show that, contrary to conventional wisdom, intra-core timing channels remain even when
using all hardware-supported flush operations. Our results furthermore
show that this is not a one-off defect of a particular implementation,
but affects multiple implementations of both
mainstream architectures.

\section{Background}\label{s:background}

\subsection{Microarchitectural State and Timing Channels}

Modern processors contain a number of microarchitectural features
that exploit temporal or spatial locality for improving average-case
performance. Inherently, these features hold state that depends on
recent execution history and affects the performance of subsequent
execution. These are the data and instruction caches, the TLB, the
branch predictor as well as code and data prefetchers. The branch
predictor typically caches state in two places, the branch target
buffer (BTB) and the branch history buffer (BHB).

On a context switch, the recent history and immediate future belong to
different processes (and thus potentially security domains), which
means that the execution of one domain can affect the timing of
the execution of another, thus establishing a channel. In fact, timing
attacks have been demonstrated via all of these microarchitectural
features, see \citet{Ge_YCH_17} for a survey.

Preventing such channels requires either avoiding any sharing of
(or contention for) microarchitectural state, or ensuring that it is
reset to a defined state (independent of execution history) on a
switch of security domain, i.e.\ flushing. Partitioning by the OS is
generally possible where resources are accessed by physical address
(which is under OS control); eg. the L2 and lower-level caches can be
partitioned by page colouring ~\citep{Kessler_Hill_92,Liedtke_HH_97,
	Shi_SCZ_11}. On-core resources, such as the L1 caches, TLB and
branch predictors, are accessed by virtual address and thus cannot be
partitioned by the OS.\footnote{The L1 caches on some processors are
	said to be physically addressed. In reality this almost always means
	that the set-selector bits are a subset of the page offset, meaning
	that the lookup only uses address bits that are invariant under
	address translation. ARM seems to be using hardware alias-detection on
	at least some cores, to make the caches be have as
	physically indexed.} In the absence of explicit hardware support
for partitioning, channels based on such resources can only be
prevented by flushing.

\subsection{Covert channels}

Where microarchitectural hardware state is not partitioned or flushed,
a Trojan can, through its own
execution, force the hardware into a particular state, and a spy can
probe this state by observing its own progress against
real time. This will constitute a covert channel, i.e.\ an information
flow bypassing the system's security policy \cite{Lampson_73}. For example, the Trojan
can modulate its cache footprint, encoding data into the number of
cache lines accessed. The spy can read the data by
observing the time taken to access each cache line. Or the Trojan can
force the branch predictor state machine into a particular state,
which the spy can sense by observing the latency of branch
instructions (i.e.\ whether they are predicted correctly).

The actual implementations of covert channels depend on the details of the
particular microarchitectural feature they exploit. A large number of
such implementations have been described, as surveyed by 
\citet{Ge_YCH_17}.More such channels have been implemented and evaluated 
	recently~\cite{Evtyushkin_PA_16,Evtyushkin_Ponomarev_16,Maurice_WSGGBRM_17}.

The threat of microarchitectural channels is not restricted to
environments compromised by Trojans.
A \emph{side channel} is a special case of a covert channel, which does
not depend on a colluding Trojan, but instead
allows a spy program to recover 
sensitive information from a non-colluding \emph{victim}. Where they exist,
side channels pose a serious threat to privacy and can be used to break
encryption~\cite{Ge_YCH_17}.

Collusion allows better utilisation of the underlying
hardware mechanism and hence covert channels tend to have much higher
bandwidth than side channels based on the same mechanism; the capacity
of the covert channel is the upper bound of the
corresponding side channel capacity. 

We focus on covert channels, as the existence of a covert channel
means that there is a potential side channel as well, particularly
where the timing channel allows the receiver to obtain address
information from the sender, as is the case in cache-based channels.
Furthermore, closing a covert channel implicitly eliminates side
channels. For that reason we focus on covert channels in this work, as
we aim to establish the degree to which microarchitectural timing
channels can be eliminated.

Historically, covert channels were mostly discussed within the scope of
multilevel security (MLS) systems~\cite{Landwehr_81, DoD_85:orange}.
Such systems have users with different classification levels and 
the system is required to ensure that a user with a high security
clearance, e.g.\ a Top Secret classification,
does not leak information to users with a lower
clearance. The security evaluation requirements for military-style
separation kernels~\cite{Rushby_81} require
evaluating covert channels and limiting their capacity \citep{US:CC:SKPP}.

The advent of modern software deployment paradigms, including cloud computing,
app stores and browsers executing downloaded code, increases the risk of covert channels.
Consequently, channels in such environment have received recent attention~\cite{Maurice_WSGGBRM_17,Lipp_GSMM_16,Wu_XW_12}.
Covert channels break the isolation guarantees of cloud environments,
where workloads of several users are deployed on the same hardware.
Similarly, mobile devices rely on the system's security
policy to ensure privacy whilst executing software of multiple, possibly
untrustworthy developers.

As indicated in the Introduction, we focus on
channels that can be exploited by a Trojan and a spy who time-share a
processor core.
This scenario implicitly eliminates \emph{transient-state}
channels~\cite{Aciicmez_Seifert_07,Yarom_GH_16},
i.e.\ those that exploit the limited bandwidth of processor
components, such as busses.
These rely on concurrent execution of Trojan
and spy, and hence do not exist in a time-sharing scenario.
We thus only need to handle 
\emph{persistent-state} channels, which rely on competing for storage
capacity of processor elements. 

\subsection{Signalling techniques}\label{p+p}
Like any communication, a covert channel requires not only a
communication medium (the microarchitectural state in our case) but also a
protocol, i.e.\ exploitation technique.

In this work we use the Prime+Probe technique
\citep{Percival_05,Osvik_ST_06}, which is commonly used
for exploiting timing channels in set-associative caching elements.
It has been
applied to the L1 D-cache~\cite{Percival_05,Osvik_ST_06},
L1 I-cache~\cite{Aciicmez_Schindler_08}, and
the LLC~\cite{Liu_YGHL_15}.
Several other techniques for exploiting cache channels have been 
suggested~\cite{Yarom_Falkner_14,Gruss_MWM_16,Gruss_MFLM_16,Osvik_ST_06},
some of which demonstrate higher capacity.
Our focus is not the maximum potential channel capacity nor the
specific mechanism used. Instead
we focus on establishing the existence of channels and whether they can be closed;
and choose Prime+Probe as a generic approach for
exploiting multiple channels.

In Prime+Probe, the spy primes the cache by filling
some of the cache sets with its own data.
The Trojan uses each of the sets that the spy primes to transmit
one bit, leaving the corresponding cache line untouched for sending a
zero, or replacing it with its own data for sending a one. 
The spy then probes the cache state to receive the information,
measuring the time taken to access the data it primed;
a long access time indicates that the Trojan replaced cache content,
representing a one, a short access time a zero.
We use a simplified implementation: the input symbol is the total
number of cache lines accessed by the Trojan (i.e.\ for simplicity we use a unary
instead of a more efficient binary encoding).

\section{Threat Model}\label{s:model}

We assume that the adversary manages to execute
a Trojan program within a restricted security domain.
For example, the adversary may compromise a service within the
restricted domain and inject the Trojan's code, or she may be a corrupt developer that
inserts malicious code into a software product used in the restricted
domain.
Executing within the restricted domain gives the Trojan access to sensitive data
which the adversary wants, however the security policies confine the
secure environment to prevent data exfiltration by the Trojan.

Additionally, the adversary controls a spy program which executes on the same
computer as the Trojan, for example, in a different virtual machine or app.
The spy is executing outside the restricted domain and consequently
can communicate freely with the adversary, but it does not have access to the sensitive data.
The adversary's aim is to exploit a microarchitectural covert channel
in the shared hardware.
If such a channel exists and is not controlled by the system, the Trojan
can use the channel to send the sensitive data to the spy,
which can then send the data to the adversary.
In this work we investigate the degree to which the system can prevent the adversary
from exploiting such covert channels.

We are focusing on time-shared use of processors, and thus ignore
transient channels.
As we are exploring microarchitectural channels,
we exclude timing channels that are controlled by software.
For example, the Trojan could create a timing channel by varying
its execution time before yielding the processor.
We note that the system can protect against such channels by
padding the execution time of the Trojan following a 
yield~\cite{Askarov_ZM_10,Cock_GMH_14,Braun_JB_15}.
Moreover, because we investigate the processor's
ability to close the channel, we only investigate
channels within the processor itself.
External channels, such as the DRAM open rows channel~\cite{Pessl_GMSM_16},
are outside the scope of this work.

\section{Methodology}\label{s:methodology}

In this work we examine the level of support
that manufacturers provide for eliminating microarchitectural
timing channels in their processors.
To this purpose, we implement multiple covert channels,
identify the processor instructions and available information 
that can be used for mitigating the channels,
and measure the capacity of the channels with and without
the mitigation techniques.
These steps are described in greater details below.

\subsection{Channels}\label{s:channels}

\citet{Cock_GMH_14} investigates efficient software mitigation
techniques against microarchitectural channels on ARM. We adopt their
techniques, viewing a channel as
a pipe into which a sender (the Trojan)  places \emph{inputs}
drawn from some set $I$ and which
a receiver (the spy) observes as \emph{outputs}
from a set $O$.
The inputs and output sets depend on the specific covert
channel used.

We implement five  channels, each targeting a different
microarchitectural component.
But we note that the microarchitectural features interact, so our
attacks cannot be orthogonal. For example, an attack targeting the
I-cache channel will implicitly probe the BTB, and thus may show a channel
even if the I-cache does not actually carry state across security domains.

Except where noted
we do not make any assumptions on the virtual-address-space
layout of either the Trojan or the spy;
the memory used for Prime+Probe attack may be allocated anywhere in the virtual address
space.

We target the channels exploiting the L1 I-cache, L1 D-cache, TLB,
branch target buffer and branch history buffer.

\noindent\textbf{L1 data cache}
For attacking the D-cache \cite{Percival_05,Osvik_ST_06}
we use the Prime+Probe implementation from the Mastik toolkit~\cite{Yarom_16_Mastik}.
%This channel uses the Prime+Probe attack technique described in
%\autoref{p+p} on the L1 D-cache~\cite{Percival_05,Osvik_ST_06}.
The input symbols enumerate the cache sets;
to send a symbol $s$,  the Trojan reads enough data to fill all ways of cache sets $0, 1, \ldots, s-1$.
The spy first fills the whole cache
with its own data, waits for a context switch (to the Trojan), and
then measures the total time taken to read a data item from each
cache set; the output symbol is the recorded time.

Note that we could use a more sophisticated encoding of the input
symbols to increase capacity. However, the point is not to establish
the maximum channel capacity, but to investigate the degree to which
it can be mitigated. We therefore keep things as simple as possible. \label{s:encoding}

\noindent\textbf{L1 instruction cache}\label{s:l1i-cache-channel}
The attack on the I-cache \citep{Aciicmez_07, Aciicmez_BG_10}
is identical to the L1 D-cache channel, except that
instead of reading data,
the programs execute code in memory locations that map to specific
cache sets. The implementation, also taken from  the Mastik code, uses
a series of jumps.

\noindent\textbf{Translation lookaside buffer}\label{s:tlb-channel}
For the TLB channel, the input set enumerates the TLB entries. The Trojan sends an
input symbol, $s$, by reading a single integer from
each of $s$ consecutive pages.
The spy measures the time to access a number of pages. In order to
reduce self-contention in the spy, it only accesses
half of the TLB.

A more sophisticated design would take into account the structure of
the TLB and aim to target the individual associative sets, and
exclude only the minimal set of pages the spy needs for its own execution.
As before, we opt for simplicity rather than capacity.

The only prior implementation of a TLB-based channel is that of \citet{Hund_WH_13},
which uses an intra-process TLB side channel to bypass
the protection of kernel address space layout randomisation (KASLR).
We are not aware of any prior implementation of inter-process TLB channels
and past work considers such channels infeasible because the
x86 TLB used to be flushed on context
switch~\cite{Zhang_Reiter_13}. However, recent x86 processors
feature a tagged TLB when operating in 64-bit mode, and ARM processor TLBs
have been tagged for a long time. As a result, TLB channels are feasible on
modern hardware.

\noindent\textbf{Branch history buffer}\label{s:branch-perdiction-unit-channel}
For exploiting the BHB we use an approach similar to the residual state-based covert channel of
\citet{Evtyushkin_PA_16:TACO}.
In each time slice, the Trojan sends a single-bit input symbol.
Trojan and spy use the same channel code for sending and receiving.
The code, the x86 version of which is shown in \autoref{f:bp-probe}, consists of a sequence of conditional
forward branches that are always taken (Line~8) which set the history to a known state.
The next code segment (Lines 10--17)
measures the time it takes to perform a branch (Line~13) that conditionally
skips over 256 \texttt{nop} instructions (Line~14).
The branch outcome depends on the least significant bit of register \texttt{\%edi}.
The return value of the code (register \texttt{\%eax}) is the measured time.
The ARM implementation is similar. 

\begin{figure}
	{\footnotesize
		\begin{tabular}{rl}
			1 & \verb|#define X_4(a) a; a; a; a| \\
			2 & \verb|#define X_16(a) X_4(X_4(a))| \\
			3 & \verb|#define X_256(a) X_16(X_16(a))| \\
			4 & \verb|| \\
			5 & \verb|#define JMP     jnc 1f; .align 16; 1:| \\
			6 & \verb|| \\
			7 & \verb|  xorl %eax, %eax| \\
			8 & \verb|  X_256(JMP)| \\
			9 & \verb|| \\
			10 & \verb|  rdtscp| \\
			11 & \verb|  movl %eax, %esi| \\
			12 & \verb|  and $1, %edi| \\
			13 & \verb|  jz  2f| \\
			14 & \verb|  X_256(nop)| \\
			15 & \verb|2:| \\
			16 & \verb|  rdtscp| \\
			17 & \verb|  subl %esi, %eax| \\
		\end{tabular}
	}
	\caption{BPU channel code for the x86 architecture.\label{f:bp-probe}}
\end{figure}

Because the code takes different paths depending on whether the input bit is set,
there is a timing difference between these two paths.
Another source of timing difference is the processor's prediction on whether the
branch in Line~13 is taken.
The channel exploits this timing difference.

To implement this channel, the Trojan and the spy map the code at the same virtual
address,
however each uses its own copy of the code
%Note that there is no need for the Trojan and the spy to use the same
(i.e.\ we do not rely on shared memory).
To send an input symbol, the Trojan sets the least significant bit of the input register
to the input symbol and repeatedly calls the code throughout the time slice.
This sets a strong prediction of the branch's outcome.
The spy, in its time slice, calls the code with the input bit cleared (branch taken)
and uses the measured time as the output symbol.

\noindent\textbf{Branch target buffer} \label{s:branch-target-buffer-channel}
To our knowledge, no BTB-based covert channel has been demonstrated to date.
To build the channel, we chain branch instructions into
a  probing buffer.  The
Trojan probes on the first \(s\) instructions, the input symbol,
while the spy measures the time taken for probing the entire buffer.
On ARM platforms, the number of branch instructions equals the known size of BTB. 
On x86 platforms, details of BTB are generally not specified by manufacturers, but can 
frequently be
reverse-engineered~\cite{Milenkovic_MK_04}. According to \citet{Godbolt_16}, the BTB 
contains 4096 entries on the Ivy Bridge microarchitecture. In our test, 
the Trojan probes from 3072 to 5120 \texttt{jmp} instructions, whereas 
the spy probes on 4096 \texttt{jmp} instructions.  The \texttt{jmp} instructions are 16-byte aligned, therefore 
the probing buffers are bigger than the L1 I-cache
(32\,KiB). 

\subsection{Mitigations}\label{s:mitigations}

The microarchitectural channels we examine in this work exploit
timing variations due to the internal state of components of the
processor core. 
The OS cannot mitigate by partitioning the components, as this is not supported by the hardware.
Furthermore, the OS cannot force partitioning through the allocation of physical addresses as the
resources are indexed virtually. The only mitigation strategy available is through hardware
cache-flushing mechanisms or completely disabling features.

\paragraph{Architectural support on  x86} \label{s:architecture-support-on-x86}

The \texttt{wbinvd} instruction invalidates all of the entries in
the processor caches~\cite{Intel_64_IA-32:asdm2_325383}.
Consequently, after the instruction executes, the caches are in a
defined state, all lines invalid, and subsequent accesses to data and
instructions are served from memory, until program locality leads to
use of recently cached data/instructions.

When executing in
32-bit mode, reloading the page-table pointer register, \texttt{CR3}, invalidates all of the entries
in the TLB and paging-structure caches except global mappings.  For
flushing the entries for global mappings, we reload the  \texttt{CR0}
register.
In 64-bit mode the TLBs are tagged with PCID, we use \texttt{invpcid}
to flush and invalidate both tagged and global TLB entries, including
paging structure caches. 

The architecture offers no instruction for flushing the state of the
prefetcher. Instead, it allows disabling the data prefetcher by
updating \texttt{MSR}~\texttt{0x1A4}~\cite{Viswanathan_14}, which
avoids any shared state in this unit. However, there is no way to
disable instruction prefetching.

In summary, on x86 we can flush all caches and the TLBs and disable the data
prefetcher. We cannot clean or disable the instruction prefetcher or
the branch predictor.

\paragraph{Architectural support on ARM} \label{s:architecture-support-on-arm} 

We use the \texttt{DCCISW} cache maintenance operation for
invalidating the data
cache and \texttt{ICIALLU} for the instruction cache. Because the L2 cache is normally
implemented as a core-external cache, operations on that cache are not
part of the ARM ISA, and we therefore use the appropriate manufacturer-specified
operations on external cache controllers to invalidate
cache ways. On the Sabre (A9) platform, we use the \emph{clean and
	invalidate set} operations. On the Hikey (A53) and TX1 (A57)
platforms, we use  \texttt{DCCISW} and \texttt{ICIALLU} for flushing
the L2 caches, which are the same operations as used for flushing the
L1 caches, but targeted at a different cache level.

On all ARM processors we use \texttt{TLBIALL} to invalidate the entire unified TLB
and \texttt{BPIALL} to invalidate all entries from branch predictors. On ARMv7, we disable
branch prediction by clearing the Z-field in the \texttt{SCTLR}
register; this operation is not supported on ARMv8. 
On the A9 we disable the L1-data prefetch by setting the
\texttt{ACTLR} register. On the A53,
we disable the data prefetcher by setting the CPU auxiliary control register.

In summary, on ARM we can flush all caches, branch prediction entries, and TLBs and disable the
data prefetcher (except on the A57). On the A9 we disable the branch
prediction.  None of the ARM platforms allow us to flush or disable the
instruction prefetcher.

\paragraph{Scrubbing state without architectural support}

Where the architecture does not support flushing or disabling a
stateful microarchitectural feature, the operating system can attempt
to programmatically reset its state during a domain switch.
For example, the context switch code can fill the L1 D-cache with
kernel data, ensuring that the contents of the cache does not
depend on prior computation.
This approach is severely limited by the insufficient documentation of
relevant microarchitectural features.
While researchers have reverse-engineered some of these features
\cite{Yarom_GLLH_15,Pessl_GMSM_15,Maurice_LNHF_15,Milenkovic_MK_04}
or the algorithm used~\cite{Gruss_MM_16,Wong_13,Bhattacharya_Mukhopadhyay_15}, 
the available information is still insufficient. Hence,  any defences that depend on such undocumented
microarchitectural properties are inherently brittle and may fail to
work as soon as a new version of the processor hits the market. We
therefore do not use such techniques, as our aim is to investigate the
degree to which microarchitectural timing channels can be reliably closed.

\subsection{Measuring the channel capacity}\label{s:capacity}
To establish whether there is a channel, we configure the Trojan
to send a pseudo-random sequence of input symbols.
The spy collects the output symbols;
a channel exists if the distribution of output symbols depends on
the input symbol. 

\begin{figure}[htb]
	\includegraphics[width=\linewidth]{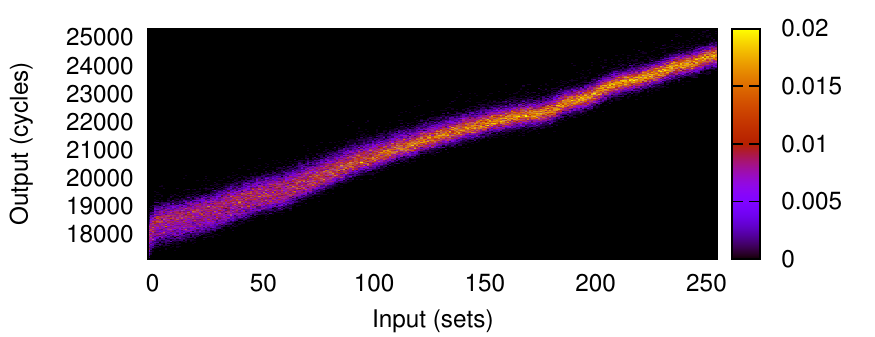}
	\caption{Channel matrix for the unmitigated L1-I covert channel on an
		ARM Cortex-A9. Colours indicate the probability of observing
		a particular output (total time in cycles for probing the buffer) for a particular input
		symbol (the number of cache sets the Trojan's code executes), as per the scale on
		the right. This example uses 909 samples for each input symbol and
		shows a channel capacity of 4~bits, we summarise this as
		\stats{909}{4.0}.
		\label{f:cm-a9-l1i-none}}
\end{figure}

\subsubsection{The channel matrix}
We use the technique of
\citet{Cock_GMH_14} to measure the information capacity of a channel.
That is, we first create a \emph{channel matrix}, which specifies
the conditional probability of an observed output symbol given an input symbol.
For example, \autoref{f:cm-a9-l1i-none} shows the
channel matrix we observe from the  L1 I-cache of an ARM Cortex-A9 processor without any countermeasures (see \autoref{s:hw} for description of
the hardware platform).
As we can see, there is a strong correlation between the input symbol
(number of cache sets that the Trojan occupies after each run)
and the output symbol (total time to jump through every cache line in  
a cache-sized buffer), resulting in a strong channel.
In the absence of a
channel the graph would show no horizontal variation.

\subsubsection{Channel capacity}
As a measure of the channel capacity,
we calculate the \emph{Shannon capacity}~\cite{Shannon_48},
denoted by $\calC$,
from the channel matrix.
$\calC$  indicates the average number of bits of information
that a computationally unbounded receiver can learn from
each input symbol.
For the channel matrix in \autoref{f:cm-a9-l1i-none} we
find that $\calC=4.0$\,b. 
Because the Trojan sends one of 257 possible values (\(0\cdots256\)),
the maximum capacity expected is theoretically $8.0$ bits per symbol.
The observed capacity is smaller than the maximum due to sampling
error, which  leads to a distribution of output values even if
there was a unique mapping from input to output symbols. Observable
capacity could also be reduced through undocumented interference by
other hardware features, such as replacement policy or prefetching.

\begin{figure}[tb]
	\includegraphics[width=\linewidth]{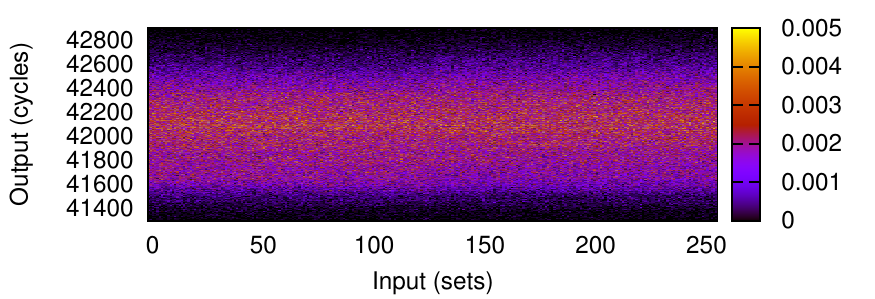}
	\caption{Channel matrix for the L1-I channel on the A9
		with cache invalidation.
		\statsm{3823}{0.70}{0.38}.
		\label{f:cm-a9-l1i-wbinvd}}\vspace{1ex}
	
	\includegraphics[width=\linewidth]{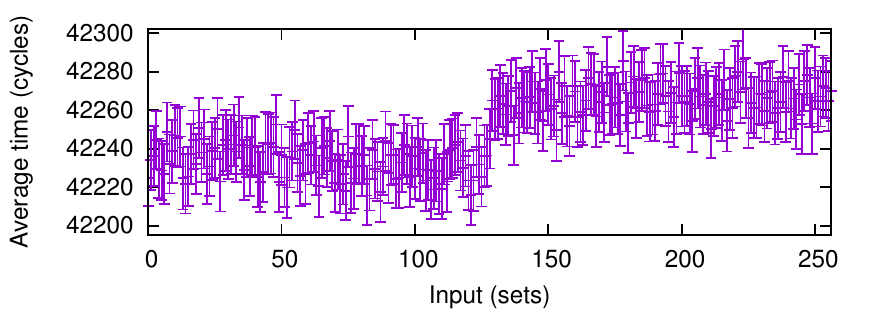}
	\caption{Average probe times, with 95\% confidence intervals, 
		for the mitigated L1-I covert channel on ARM Cortex-A9.
		\label{f:avg-a9-l1i-wbinvd}}
\end{figure}

\begin{table*}[tb]\center
	\caption{Experimental hardware, covering 2--3 generations of
		microarchitectures across x86 and ARM. I-TLB and D-TLB represent first-level TLBs, L2~TLB  represents the
		second-level (unified) TLB. ``?'' indicates unknown values.   \label{t:archs}
	}
	\footnotesize 
	\begin{tabular}{ll|rrrrrr}
		\multicolumn{2}{l|}{Architecture} &\multicolumn{1}{c}{\textbf{x86}} & \multicolumn{1}{c}{\textbf{x86}} &  \multicolumn{1}{c}{\textbf{x86}} &
		\multicolumn{1}{c}{\textbf{ARMv7}} & \multicolumn{1}{c}{\textbf{ARMv8}} & \multicolumn{1}{c}{\textbf{ARMv8}} \\
		\multicolumn{2}{l|}{Microarchitecture}  &\multicolumn{1}{c}{\textbf{Sandy Br}} & \multicolumn{1}{c}{\textbf{Haswell}} & 
		\multicolumn{1}{c}{\textbf{Skylake}} & 
		\multicolumn{1}{c}{\textbf{A9}} & \multicolumn{1}{c}{\textbf{A53}} & \multicolumn{1}{c}{\textbf{A57}}\\ \hline
		\multicolumn{2}{l|}{Manufacturer} & Intel & Intel  & Intel  & Freescale & HiSilicon & NVIDIA \\
		\multicolumn{2}{l|}{Processor} & i7-2600 &  E3-1220 v3 & i7-6700 & i.MX6 & Kirin 620 & Jetson TX1 \\ \hline
		\multicolumn{2}{l|}{Clock rate (GHz)} & 3.4 & 3.1 & 3.4 & 0.8 & 1.2  & 1.91  \\
		\multicolumn{2}{l|}{Year} & 2012 & 2013 & 2015 & 2012 & 2014 & 2015 \\
		\multicolumn{2}{l|}{Address size (bit)} & 32 or 64 & 32 or 64 & 32 or 64 & 32 & 32 or 64 & 32 or 64\\
		\multicolumn{2}{l|}{Execution order} & OoO & OoO & OoO & OoO & InO & OoO \\\hline
		\multicolumn{2}{l|}{Time slice (ms)} & 1 & 1 & 1 & 2 & 3 & 3 \\\hline
		
		L1-D & size (KiB) & 32 & 32 & 32 & 32 & 32 & 32  \\
		& line (B) & 64 & 64  & 64 & 32 & 64 & 64  \\
		& sets$\times$assoc. & 64$\times$8 & 64$\times$8 & 64$\times$8 & 256$\times$4 & 128$\times$4 & 256$\times$2 \\
		%& assoc. & 8 & 8 & 8 & 4 & 4  & 2 \\ 
		\hline
		
		L1-I & size (KiB) & 32 & 32 & 32 & 32& 32 & 48 \\
		& line (B) & 64 & 64 & 64 & 32 & 64 & 64 \\
		& sets$\times$assoc. & 64$\times$8 & 64$\times$8 & 64$\times$8 & 256$\times$4 & 256$\times$2 & 256$\times$3 \\
		%& assoc. & 8 & 8 & 8 & 4 & 2 & 3 \\\hline
		\hline
		
		L2   & size (KiB) & 256 & 256 & 256  & 1024 & 512  & 2048 \\ % $\times2$\\
		& line (B) & 64 & 64 & 64 & 32 & 64 & 64 \\
		& sets$\times$assoc. & 512$\times$8 & 512$\times$8 & 512$\times$8 & 2048$\times$16 & 512$\times$16 &  2048$\times$16 \\
		%& assoc. & 8 & 8 & 8 & 16 & 16 & 16 \\\hline
		\hline
		
		L3   & size (MiB) & 8 & 8  & 8 & N$\times$A & N$\times$A & N$\times$A  \\
		& line (B) & 64  & 64  & 64  & N$\times$A & N$\times$A & N$\times$A\\
		& sets$\times$assoc. & 8192$\times$16 & 8192$\times$16 &  8192$\times$16 &  N$\times$A& N$\times$A  & N$\times$A \\
		%     & assoc. & 16 & 16 & 16  & N$\times$A & N$\times$A & N$\times$A  \\\hline
		\hline
		
		BTB  & size & ? & ? & ? & 512 & 256  & 256  \\
		& sets$\times$assoc. & ? & ? & ? &  256$\times$2 & ?  & ?  \\
		%     & assoc. & ? & ? & ? & 2 & ? & ?  \\\hline
		\hline
		
		I-TLB & size & 128 & 128 & 128 & 32 & 10 & 48 \\
		& sets$\times$assoc. & 32$\times$4 & 16$\times$8 & 16$\times$8 & 32$\times$1 & 10$\times$1 & 48$\times$1 \\
		%    & assoc. & 4 & 8 & 8 &  1 & 1 & 1 \\ \hline
		\hline
		
		D-TLB & size & 64 & 64 & 64 & 32 & 10 & 32 \\
		& sets$\times$assoc. & 16$\times$4 & 16$\times$4  & 16$\times$4 & 32$\times$1 &  10$\times$1 & 32$\times$1 \\
		%    & assoc. & 4 & 4 & 4 & 1 &  1 & 1 \\\hline
		\hline
		
		L2~TLB & size & 512 & 1024 & 1536 & 128 & 512 & 1024 \\
		& sets$\times$assoc. & 128$\times$4 & 128$\times$8 & 128$\times$12 & 64$\times$2& 128$\times$4 & 256$\times$4 \\
		%    & assoc. & 4 & 8 & 12 & 2 & 4 & 4 \\
		\hline
	\end{tabular}
\end{table*}

\subsubsection{Channel bandwidth}
The channel bandwidth can now be calculated by
multiplying the capacity by the input symbol rate, i.e.\ the frequency
with which the Trojan can send different symbols. As the processor is
time-shared between the Trojan and the spy, the Trojan can send at
most one symbol per time slice. In the case of
\autoref{f:cm-a9-l1i-none} we use a time slice of 1\,ms, meaning that
per second the Trojan gets 500 opportunities to send a symbol, as the
two processes run in alternating time slices.
The potential bandwidth of the channel is therefore
$500\times 4=2,000$\,b/s. We use different time slices on the various
platforms, as specified in \autoref{t:archs}.

Much higher bandwidths have been demonstrated in the
literature~\cite{Liu_YGHL_15,Gruss_MWM_16,Evtyushkin_Ponomarev_16,Maurice_WSGGBRM_17}.
One reason for our relatively low bandwidth is that our channel is between
programs that time-share a processor, whereas the other channels
were observed between concurrently executing processes,
allowing for a much higher symbol transfer rate. Furthermore, as
indicated in \autoref{s:encoding}, we use an unsophisticated encoding,
where the input is simply the number of cache sets accessed,
resulting in 257 possible values. Instead we could map each bit of the
input symbol's binary representation onto a cache set, resulting in 
\(2^{256}\) values and a theoretical maximum capacity of 256~bits per
symbol (although the actually achievable capacity would be much
smaller, given the spread of the distribution). 
Instead of aiming for maximum capacity, our channels are designed
to facilitate analysis.

\subsubsection{Visualising low-capacity channels}
Visualising the channel matrix as a heat map is useful for presenting the
wealth of information that the matrix contains.
However, for low-capacity channels, such a representation may fail to
highlight the channel. For example, when mitigating the L1-I-cache
channel on the A9 using  methods for flushing and invalidating all caches (\autoref{s:architecture-support-on-arm}), we obtain the
channel matrix of \autoref{f:cm-a9-l1i-wbinvd}. There is no channel
evident, but the picture is fuzzy enough to hide a small
channel of 0.7\,b, about a sixth of the original capacity.

For better visualisation 
we compress the information of the channel matrix by averaging the
output symbols for each input set, which yields the graph of
\autoref{f:avg-a9-l1i-wbinvd}. We now clearly see the residual
channel, with a clear difference between the average output
symbol for small vs.\ large input values.

\subsubsection{Verifying low-capacity channels}
A challenge for our channel capacity measurement
is that it is based on a sampled distribution.
Sampling errors may result in an apparent non-zero channel,
even when in reality  there is no channel.
Consequently, when the computed channel
capacity $\calC$ is low,
we need to determine whether we observe  a low-capacity
channel or just sampling noise.
For example,  while \autoref{f:avg-a9-l1i-wbinvd} clearly shows that
there is a residual channel, even in the averaged graph it might not
be obvious whether there is a channel or just sampling noise; we need
a systematic way of detecting a statistically significant channel
without relying on visual inspection. Specifically,
we need to distinguish between the null hypothesis,
that states that there is no channel,
i.e.\ the distribution of output symbols is independent
of the input symbols,
and the 
alternative hypothesis that there is a channel.

Following \citet{Cock_GMH_14}, we bound the effect of
sampling error by simulated sampling of a true zero-capacity
channel. Specifically, we
randomly distribute the output symbols collected
across all input symbols and measure the capacity of this
simulated channel.
We repeat the process 1000 times, generating 1000 simulated capacities.
If our measured sample (\autoref{f:avg-a9-l1i-wbinvd}) is drawn from a
single distribution (i.e.\ no channel),
there is a high likelihood that some of the simulated capacities would
be higher than the observed $\calC$.
Thus, if $\calC$ is bigger than the maximum, $\Cmax$, of the 1000 simulated
capacities, the probability that the sample is drawn from
a single distribution is less than 0.1\% and we reject the null hypothesis.
Otherwise, we conclude that we cannot reject the null hypothesis
and that the test is inconclusive (i.e.\ consistent with no channel).

In the case of \autoref{f:cm-a9-l1i-wbinvd}, we find that of the observed capacity
\(\calC=0.7\)\,b, at most \(\Cmax=0.38\)\,b can be attributed to
sampling error, meaning that we have a true residual channel of at
least \(\calC=0.3\,b\). Given the 1\,kHz context-switching rate, this
implies a residual bandwidth of at least 150\,b/s, enough to leak a
typical RSA key in less than half a minute.

\newcommand{\NM}{\multicolumn{2}{c}{N/M}}
\newcommand{\NA}{\multicolumn{2}{c|}{N/A}}
\newcommand{\NAl}{\multicolumn{2}{c}{N/A}}
\newcommand{\Rw}[1]{\multicolumn{1}{r@{}}{#1}&\multicolumn{1}{l|}{}}
\newcommand{\Rwl}[1]{\multicolumn{1}{r@{}}{#1}&\multicolumn{1}{l}{}}
\begin{table*}[tb]\center
	\caption{Observed unmitigated (``none'') and maximally mitigated (``full'')
		channel capacities \(\calC\) and maximum apparent capacities for
		empty channel
		\(\Cmax\) in bits.  Channels that cannot be mitigated with all hardware-provided operations are marked in red. \label{t:results}}
	\footnotesize
	\begin{tabular}{ ll| r@{}>{~(}l<{)} | r@{}>{~(}l<{)} | r@{}>{~(}l<{)} | r@{}>{~(}l<{)} | r@{}>{~(}l<{)} | r@{}>{~(}l<{)} | r@{}>{~(}l<{)} }
		\multicolumn{2}{r|}{\bf Processor} & \multicolumn{2}{c|}{\textbf{Sandy Bridge}} & \multicolumn{4}{c|}{\textbf{Haswell}} 
		& \multicolumn{2}{c|}{\textbf{Skylake}}
		& \multicolumn{2}{c|}{\textbf{A9}} & \multicolumn{2}{c|}{\textbf{A53}}
		& \multicolumn{2}{c}{\textbf{A57}} \\
		\bf Chan.\hspace*{-1em}& \bf Mitig.& \multicolumn{2}{c|}{32-bit} & \multicolumn{2}{c|}{32-bit} & \multicolumn{2}{c|}{64-bit} &  \multicolumn{2}{c|}{64-bit} & \multicolumn{2}{c|}{32-bit} & \multicolumn{2}{c|}{32-bit} & \multicolumn{2}{c}{32-bit} \\
		\hline \hline
		
		\multirow{2}{*}{L1-D}
		&none& \Rw{4.0} & \Rw{4.1} & \Rw{4.7} & \Rw{3.3} & \Rw{5.0} & \Rw{2.8} & \NM \\
		&full& 0.038&0.025 & 0.083&0.050 &  0.43&0.24 & 0.18&0.01 
		&  0.11&0.046  & \textcolor{red}{0.15}&\textcolor{red}{0.079} &  \NM \\
		\hline

		\multirow{2}{*}{L1-I}
		&none& \Rw{3.7} & \Rw{0.65} & \Rw{0.46} & \Rw{0.37} & \Rw{4.0} & \Rw{4.5} & \Rwl{6} \\
		&full&  \textcolor{red}{0.85}& \textcolor{red}{0.05}  
		&  \textcolor{red}{0.25}& \textcolor{red}{0.04} 
		&  \textcolor{red}{0.36}& \textcolor{red}{0.1}
		&  \textcolor{red}{0.18}&\textcolor{red}{0.09}
		&  \textcolor{red}{1.0}& \textcolor{red}{0.72}    
		&  \textcolor{red}{0.5}& \textcolor{red}{0.24} 
		&  \textcolor{red}{2.8}& \textcolor{red}{0.34}  \\
		\hline
		
		\multirow{2}{*}{TLB} 
		&none&  \Rw{3.2}   & \Rw{3.1} & \Rw{3.2} & \Rw{2.5}  & \Rw{0.33}  & \Rw{3.4}  & \NM \\
		&full&    0.47&0.29    &   0.37&0.23   
		& 0.18&0.1 & 0.11&0.06 
		& 0.16&0.087  & 0.14&0.08  & \NM \\
		\hline
		
		\multirow{2}{*}{BHB}
		&none& \Rw{1.0} & \Rw{1.0} & \Rw{1.0} & \Rw{1.0} & \Rw{1.0}  & \Rw{1.0}  & \NM \\
		&full&  \textcolor{red}{1.0}&\textcolor{red}{0.023}        &  \textcolor{red}{1.0}&\textcolor{red}{0.009}  &
		\textcolor{red}{1.0}&\textcolor{red}{0.002}        &  \textcolor{red}{1.0}&\textcolor{red}{0.002}  
		& 0.009&0.008  &  \textcolor{red}{0.5}& \textcolor{red}{0.02}   & \multicolumn{2}{c}{ N/M}  \\
		\hline
		
		\multirow{2}{*}{BTB}
		&none&  \Rw{2.0} & \Rw{4.6} &  \Rw{4.1} &  \Rw{1.8} & \Rw{1.1} & \Rw{1.3} & \NM \\
		&full& \textcolor{red}{1.7}&\textcolor{red}{1.2} & \textcolor{red}{1.3}&\textcolor{red}{0.4} & \textcolor{red}{1.6}&\textcolor{red}{0.5} & \textcolor{red}{1.9}&\textcolor{red}{1.2} 
		& 0.068&0.032 & \textcolor{red}{0.15}&\textcolor{red}{0.08} &\NM\\
		\hline
	\end{tabular}
\end{table*}

\section{Results}\label{s:results}

\subsection{Evaluation platforms}\label{s:hw}

We examine processors from the two most widely-used architectures,
x86 and ARM. For x86 these are the Sandy Bridge, Haswell and Skylake
microarchitectures. For ARM we use a Cortex-A9, an implementation of ARMv7, and
a Cortex-A53 and A57, in-order (InO) and  out-of-order (OoO)
implementations of ARMv8, the latest version of
the architecture. We summarise their relevant features in
\autoref{t:archs}. Note that the information on the branch predictors
are incomplete, as there is very limited information available. 

As we are probing hardware properties, the results should be
independent of the OS or hypervisor used. However, the more complex
the code base, the more complicated and error-prone it is to implement
mitigations. Furthermore, with a large system, such as Linux or Xen,
it becomes harder to preclude interference from various software
components. We therefore use the seL4
microkernel~\cite{seL4,Klein_AEMSKH_14}, which is a small (about 10,000 lines of code)
and simple system. Furthermore, seL4's comprehensive formal
verification includes
proof of freedom from storage channels~\cite{Murray_MBGBSLGK_13}, which means
any remaining channels must be timing channels, simplifying analysis
of results.

Specifically, we use the separation
kernel~\cite{Rushby_84} configuration of seL4, where we run our Trojan
and spy code ``bare metal'', with no actual OS services. This means
that we are working in an almost
noise-free environment, which helps measuring small-capacity channels.

\subsection{Overview of results}\label{s:results-sum}

\paragraph{Raw channels}

The ``none'' rows in \autoref{t:results} show the unmitigated 
channel capacities of all the channels we test across our six processors.
We see that most of the channels leak several bits per
input symbol. The main exceptions are the I-cache channel on the
recent x86 processors and the TLB channel on the A9, which have
capacities in the range of 0.33--0.65 bits. We do not know the reasons
for these lower capacities, but suspect that replacement policies
might reduce the capacity seen by our simple attack, which works best
with true LRU. A more sophisticated implementation, taking into
account the (reverse-engineered) replacement policy, might show a
bigger channel.

The main take-away from these results is that \textbf{all} microarchitectural
features that cache recent execution history can be used for timing
channels. This unsurprising result confirms that all such features
must be taken into account when trying to stop timing-channel leakage.

\paragraph{Mitigated channels}

The ``full'' rows in \autoref{t:results} show the observed channel
capacities with all mitigations enabled. This goes beyond just
flushing caches, but using all mitigations discussed in
\autoref{s:mitigations}; we will discuss specific examples below.

From the table
we make the surprising observation that \textbf{none of the channels
	are completely closed by employing all available hardware functions
	for removing the microarchitectural state}. All observed channels are larger than the
statistical bound allowed by the null hypothesis.
In fact, for each processor we find at least one channel that leaks over half a bit
per symbol (which translates into a bandwidth of hundreds of bits per second).

While all residual channels are statistically significant, we
highlight (in \textcolor{red}{red font}) the cases where the observed
capacity is well above the sampling accuracy or where we can see from the
visual representation that there is a definite channel.

In some cases the explanation for the remaining channels is straightforward.
For example, Intel architectures do not support any method of clearing the
state of the BPU. 
Consequently, the branch prediction channel remains unchanged
even when we enable all of the protection provided by the processor (\autoref{s:d-branch-channel-result}).
In other cases the story is more intricate.  We will explore some of
them in more depth.

\subsection{Finding 1: I-cache channel persists on all platforms}\label{s:discovery1}

In particular, the L1-I cache channel remains even with all mitigation method enabled on 
all the testing platforms. This seems particularly surprising, as one
would expect that a full cache flush should be sufficient to remove
this channel. We pick one microarchitecture from each ISA to explore further.

\subsubsection{A9 I-cache channel}
We have already done a partial examination of the A9's I-cache
channel, when we used it as an example in \autoref{s:capacity}.
Recall that \autoref {f:cm-a9-l1i-none} showed a clear linear
correlation between input and output symbols. Recall further that a
clear channel remained when invalidating the cache on context
switches, as evidenced by the clear transition between two distinct
distribuions in \autoref{f:avg-a9-l1i-wbinvd}.

On the A9, the distance between two addresses that map to the same cache set,
known as the cache \emph{stride} (and equal to the cache size divided
by associativity),  is 8\,KiB. 
Clearly, the  transition in \autoref{f:avg-a9-l1i-wbinvd}
occurs at 4\,KiB, which matches the page size.
This may indicate that the channel originates not from the cache itself,
but from some part of the virtual memory unit,
for example, from the TLB.
Hence, clearing the TLB might eliminate this channel.

\begin{figure}[htb]
	\includegraphics[width=\linewidth]{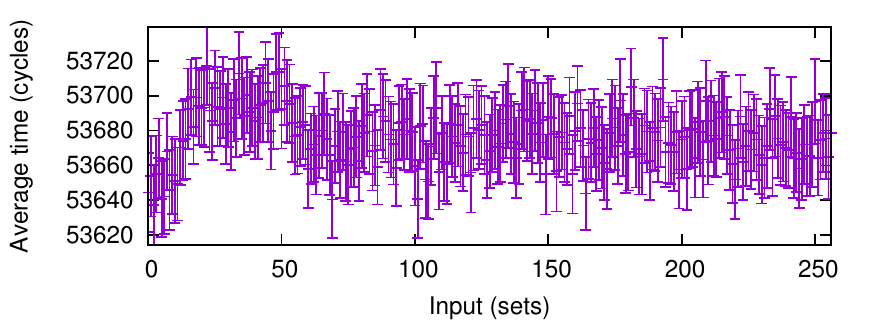}
	\caption{Average probe times, with 95\% confidence intervals, for the L1-I 
		I covert channel on A9 with all mitigations.
		\statsm{909}{1.0}{0.72}.
		\label{f:avg-a9-l1i-all}}
\end{figure}

We next apply all of the countermeasures available on the A9,
including TLB flush, which yields the average probing cost shown in
\autoref{f:avg-a9-l1i-all}. Clearly, despite using everything the
hardware gives us, there is still a significant channel:
Input values smaller than 14 result in below-average 
output symbols, whereas symbols in the range 15-50 produce above-average
output.

One possible explanation for the channel is that
the processor maintains some state that is not cleared
by flushing the caches and is not deactivated when disabling the data prefetcher.
An alternative explanation is that the state is maintained outside the processor,
for example resulting from a DRAM open-row channel~\cite{Pessl_GMSM_16}.

\begin{figure}[htb]
	\includegraphics[width=\linewidth]{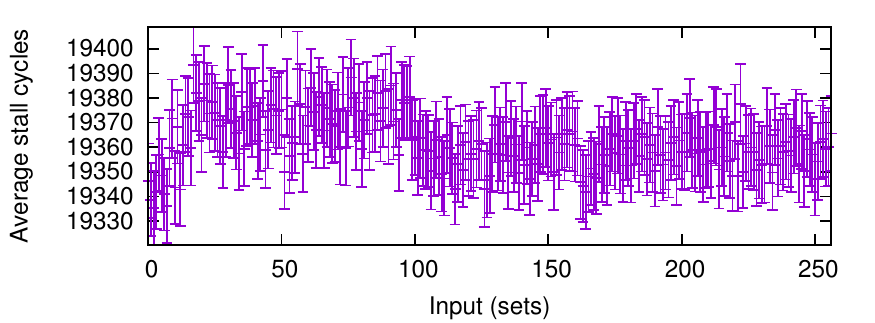}
	\caption{Average instruction cache dependent stall cycles with 95\%
		confidence intervals for  A9 with all mitigations.
		\label{f:avg-a9-l1i-all-pmu}}
\end{figure}

To investigate further, we use the
performance monitoring unit (PMU) to count the instruction-cache-dependent stall cycles (\autoref{f:avg-a9-l1i-all-pmu}). 
We observe that the stall cycles show the same trend for small input
values as the probe cost of \autoref{f:avg-a9-l1i-all}. As these
instruction dependent stalls are triggered by internal
processor state, we can conclude that the channel is not (solely) caused by
core-external features.

\paragraph{A53}
We similarly use the PMU to analyse the cause of the residual timing channel on the A53. 
We find that the number of L1 I-cache refills increases as the probing
timing decreases (graph omitted for space reasons).

A possible explanation is that the state of instruction prefetcher,
which cannot be disabled, might be affected by the Trojan. Because there is no official 
documentation on the implementation of the instruction prefetcher, we cannot
investigate further.

\subsubsection{Sandy Bridge I-cache channel}\label{s:sandy-l1i-cache-channel}

\begin{figure}[htb]
	\includegraphics[width=\linewidth]{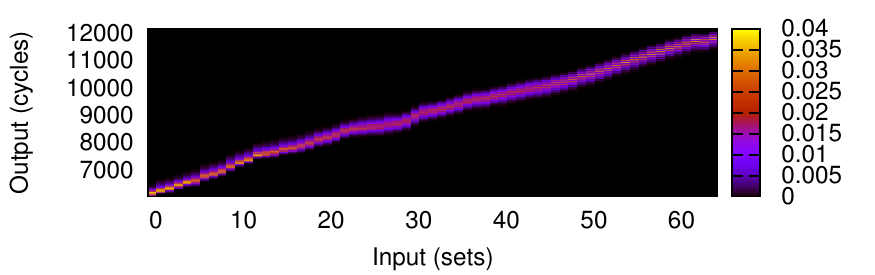}
	\caption{Channel matrix for the L1-I covert channel on Sandy Bridge
		without countermeasures.
		\stats{15420}{3.7}.
		\label{f:cm-sb-l1i-none}}
\end{figure}

\begin{figure}[htb]
	\includegraphics[width=\linewidth]{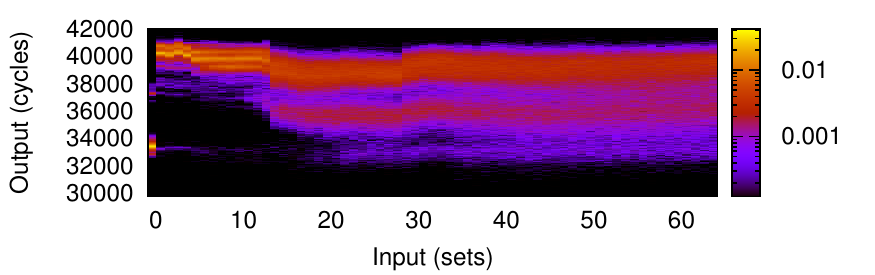}
	\caption{Channel matrix for the L1-I covert channel on Intel Sandy Bridge
		with cache invalidation.
		\stats{15415}{1.4}.
		\label{f:cm-sb-l1i-wbinvd}}
\end{figure}

\autoref{f:cm-sb-l1i-none} shows that without countermeasure this
channel behaves like expected (and qualitatively similar to the
A9). Again, flushing the cache is only moderately effective, reducing
the capacity by not even 65\%, although the channel matrix looks quite
different, as shown in \autoref{f:cm-sb-l1i-wbinvd}.

\begin{figure}[htb]
	\includegraphics[width=\linewidth]{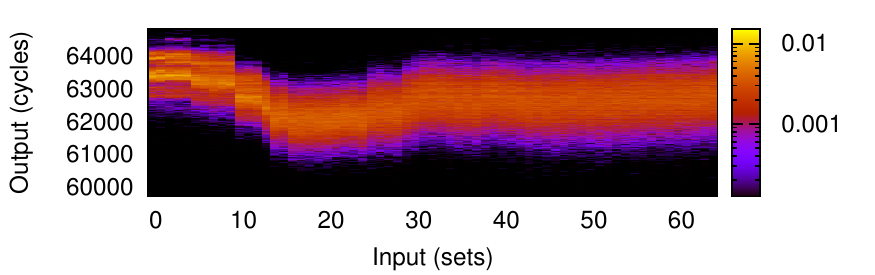}
	\caption{Channel matrix for the L1-I channel on Sandy Bridge
		with all mitigations.
		\statsm{15415}{0.85}{0.05}.
		\label{f:cm-sb-l1i-all}}
\end{figure}

If we apply our remaining countermeasures, flushing the TLB and
disabling the data prefetcher, a distinct channel remains, with a capacity of
just under a quarter of the original, as
shown in \autoref{f:cm-sb-l1i-all}. The PMU shows that the number of instruction cache, streaming buffer and victim cache misses
(ICACHE.MISSES) increases as the probing time decreases. Again, this
rules out core-external effects and makes us suspect contention
between  Trojan and spy on some unknown instruction prefetcher.

A potential alternative explanation is that the cache-flush operation does
not operate as advertised. As instructions are normally read only, the
I-cache is normally coherent with memory and does not need
flushing. Could it be that the hardware is taking a shortcut and does
not actually invalidate the I-cache?

To test this hypothesis we change our exploit code to re-write the
probing buffer (a chain of jumps) before flushing the cache  (using
\texttt{clflush}). This makes the I-cache incoherent with memory,
leaving the hardware no choice but perform the actual flush
when requested. However, 
the residual timing channel remains on both 64-bit Haswell
($\calC=0.2$\,b, \(\Cmax=0.03\)\,b) and Skylake machines ($\calC=0.2$\,b, \(\Cmax=0.06\)\,b); graphs not
shown for space reasons. 

\subsection{Finding 2: BTB  channel cannot be mitigated on x86} \label{s:btb-timing-channel-x86}

The I-cache results in \autoref{t:results} were obtained in a setup
where the Trojan and spy had their buffers of chained \texttt{jmp}
instructions allocated at the same virtual address. As far as the
I-cache is concerned, the actual virtual addresses should not matter,
as long as the buffers are contiguous in the virtual address space.

\begin{figure}[htb]
	\includegraphics[width=\linewidth]{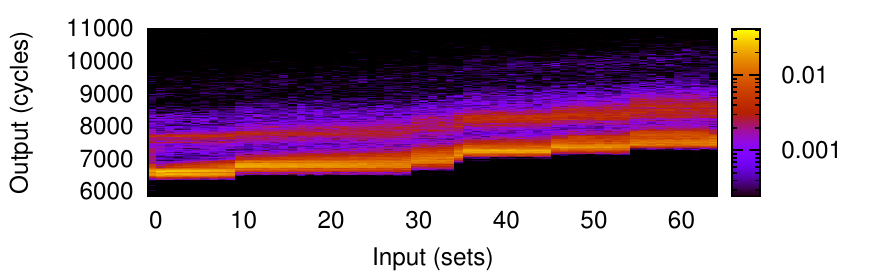}
	\caption{Channel matrix for the L1-I channel using different virtual addresses on Skylake
		with all mitigations.
		\statsm{7438}{1.2}{0.2}.
		\label{f:cm-sk-l1i-all-diff-vir}}
\end{figure}

To investigate further, we change the setup to use different virtual
address ranges for the two buffers (0x2c94000 in the Trojan, 0x2c82000
in the spy), resulting in a \emph{stronger} channel, as visualised in
\autoref{f:cm-sk-l1i-all-diff-vir}. This could be an effect caused by
contention in the branch predictor, or some prefetcher, as discussed
in \autoref{s:sandy-l1i-cache-channel}.

\begin{figure}[htb]
	\includegraphics[width=\linewidth]{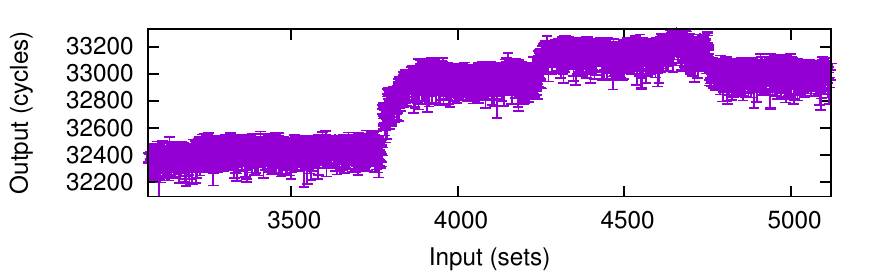}
	\caption{Average probing times, with 95\% confidence intervals
		for the BTB channel on 64-bit Haswell
		with all mitigations.
		\statsm{893}{1.6}{0.5}.
		\label{f:avg-hw-btb-wbinvd}}
\end{figure}

In fact, \autoref{t:results} shows clear BTB channels on all
platforms. They are particularly strong on the x86 processors, despite
this being a black-box attack, given the complete lack of
documentation of the x86 BTBs. An attack based on some understanding
of the BTB implementations would likely show even bigger
channels. Also, there is no architectural support for flushing these
channels. \autoref{f:avg-hw-btb-wbinvd} shows an example (on Haswell)
of this very definite channel with all mitigations deployed.

\subsection{Finding 3: Branch-history-buffer channel persists} \label{s:d-branch-channel-result}

\begin{figure}[htb]
	\includegraphics[width=\linewidth]{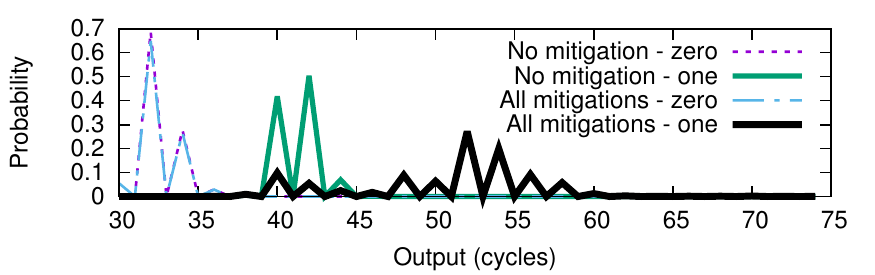}
	\caption{Distribution of output values for input values
		in the branch prediction channel on Skylake,
		with and without mitigations. The zero curves are indistinguishable.
		\label{f:bpu-skylake}} 
\end{figure}

We now turn our attention to the BHB channel.
Recall from \autoref{s:branch-perdiction-unit-channel} that the
this channel
only has two input values,~0 and~1, corresponding to
a branch taken and not taken. Clearly, the maximun possible capacity
of this channel is 1\,b.

\subsubsection{Skylake BHB channel}
\autoref{f:bpu-skylake} shows the distribution of output values for
these inputs without and with mitigations on Skylake.

We first observe that for both cases, the distribution of the 
output symbols for inputs 0 and~1 are clearly distinct.
For the non-mitigated case, the median output value for input~0 is 32~cycles,
whereas the median output for input~1 is~42.
Applying all 
mitigation (none of which target the BHB specifically, as the
architecture provides no mechanism) changes access times.
However, the output values for inputs~0 and~1 are now even more separated than
in the case of no mitigation, with median output values being 32
and~52, respectively. In short, our mitigation attempts are completely
ineffective on the BHB channel.

\subsubsection{A53 BHB channel} 

\begin{figure}[htb]
	\includegraphics[width=\linewidth]{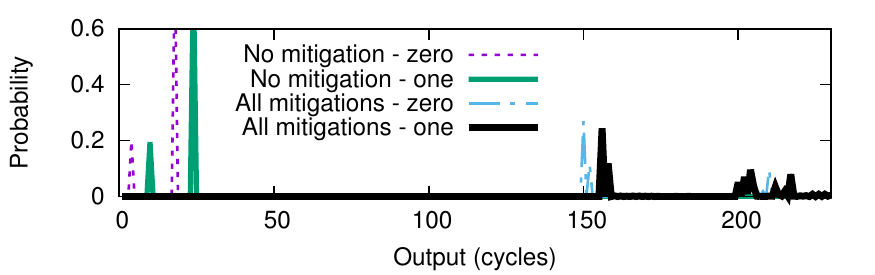}
	\caption{Distribution of output values for input values
		in the BHB channel on the A53,
		with and without mitigations.
		\label{f:bpu-hikey}} 
\end{figure}

\autoref{f:bpu-hikey} shows for the A53 that without mitigation, the
median outputs are 18~cycles and 24~cycles for inputs~0 and~1,
respectively. With mitigation,
including invalidating branch prediction entries (\texttt{BPIALL}),
the median outputs are 202~cycles for both inputs, but the
well-defined minima of the distributions are 150 vs.\ 156~cycles,
producing a clear residual channel.

\subsection{Finding 4: TLB flushing is insufficient for removing the x86 TLB channel}\label{s:discovery4}

\begin{figure}[htb]
	\includegraphics[width=\linewidth]{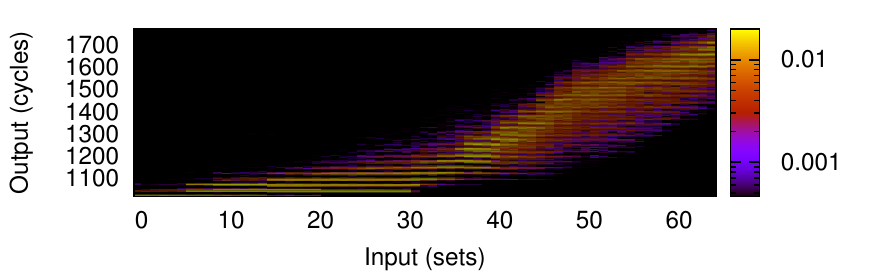}
	\caption{Channel matrix for the TLB covert channel on  Skylake
		with TLB flushing.
		\stats{3818}{2.2}.
		\label{f:cm-sl-tlb-tlb}}
\end{figure}

We finally turn our attention to the TLB, which, according to
\autoref{t:results}, has a
very distinct channel (2.5~bits). Because the TLB entries are tagged with address space ID, the 
timing reveals the L1 TLB contention. For testing the effectiveness of
TLB flush by using \texttt{invpcid}, we explicitly
flush all the TLB entries including paging caches.  As
\autoref{f:cm-sl-tlb-tlb} shows, a strong channel remains, in fact,
the capacity is almost unaffected by the flush.

\begin{figure}[htb]
	\includegraphics[width=\linewidth]{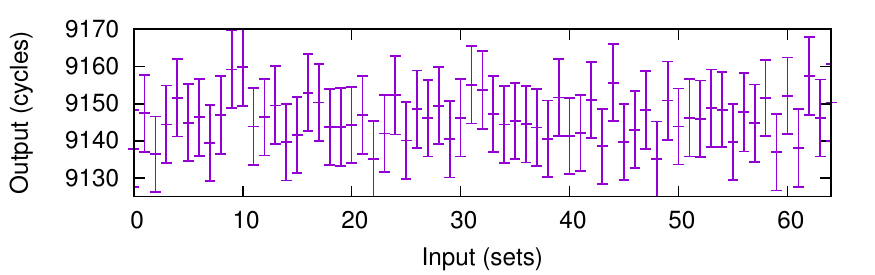}
	\caption{Average probing times, with 95\% confidence intervals  for the TLB covert channel on Skylake 
		with all mitigations.
		\statsm{19335}{0.11}{0.06}.
		\label{f:avg-sl-tlb-wbinvd}}
\end{figure}

The reason is presumably that the x86 architecture caches TLB entries
in the data cache.
When applying all mitigations listed in \autoref{s:architecture-support-on-x86}, the channel is mostly closed
as shown in \autoref{f:avg-sl-tlb-wbinvd}. While \(\calC\) is still
about twice \(\Cmax\), both are quite small, and
from \autoref{f:avg-sl-tlb-wbinvd} it seems that the remnant channel
would be hard to exploit.

\section{Discussion}\label{s:discussion}

Our results clearly show that many microarchitectural features aimed at
improving average-case performance produce high-capacity timing channels
that cannot be closed with any available countermeasures, including
those often suggested
for mitigating the exact channels we explore
\cite{Osvik_ST_06,Godfrey_Zulkernine_13,Zhang_Reiter_13}, and despite some
of them being prohibitively expensive \cite{Guanciale_NBD_16,Godfrey_Zulkernine_13}.
As we indicated earlier,
we do not build the channels for maximum capacity, and
further engineering is likely result in even higher capacities. Moreover, for high-security systems,
even channels with capacities below one bit per second
may pose a threat.
For example, the Orange Book~\cite{DoD_85:orange} recommends
that channels with a bandwidth above 0.1~bits per
second are audited. 

The root of the problem is that many microarchitectural features
either explicitly cache recent execution history (e.g.\ branch target
buffers), or accumulate state based on such history
(prefetcher state machines). Whenever such state is preserved across a context
switch, a timing channel will result.

The OS has no mechanisms to mitigate such channels without hardware
support. Specifically, to prevent timing channels, the OS must be
given the opportunity to either cleanly partition such state between
security domains, or flush on every switch of security
domain. Partitioning is easily achieved with large physically-addressed
caches through page colouring \citep{Kessler_Hill_92,Liedtke_HH_97,
	Shi_SCZ_11}, but is probably not feasible for the
virtually-addressed on-core resources. Hence, \emph{there must be
	architected mechanisms for flushing all history-dependent on-core
	state}. 

Obviously, such mechanisms should not affect performance where they
are not needed, and the performance impact should be minimised where
they are deployed. For example, the x86 architecture provides no
architected mechanism for selectively flushing the L1 caches, the only
cache-flush operation, \texttt{wbinvd}, flushes the complete cache
hierarchy. From the security point of view, this is complete overkill,
as the OS can easily partition caches other than the L1. In contrast,
flushing just the L1 caches on a partition switch, i.e.\ at a rate of
no more than 1000\,Hz, will have no appreciable performance impact, as
the direct and indirect costs of the flush should be in the
microsecond range, and no L1 content is likely to be hot after a
context switch.

Similarly, direct and indirect cost of flushing branch predictors and
prefetchers should be negligible, provided efficient hardware support
is available. 

One way to look at our results is to argue that the ISA, the
traditional hardware-software contract, is insufficient for building
truly secure systems. \emph{The ISA is sufficient to build software
	that is functionally correct but it is insufficient for building
	software that is secure, i.e.\ able to preserve confidentiality in the
	presence of untrusted code.} Arguably, for security we need a new
hardware-software contract that contains enough information about
microarchitecture to ensure secure partitioning or time-sharing.

\section{Conclusions}
We investigated intra-core covert timing channels in multiple
generations of x86 and ARM processors, and found that all investigated
platforms exhibited high-capacity channels that cannot be closed with
any known mechanism, irrespective of cost.

We therefore have to conclude that \emph{modern mainstream processors
	are not suitable for security-critical uses where confidentiality
	must be preserved on a processor core that
	is time-multiplexed between different security domains}.

This work only explores the tip of the iceberg.
We have limited ourselves to intra-core channels in a time-sharing scenario.
In doing that we ignored all transient-state covert channels attacks and
all attacks that rely on state outside the processor. Furthermore, the
hardware contention caused by shared buses remain a serious security risk
for threads sharing a platform.

The inevitable conclusion is that \emph{security is a losing game until the
	hardware manufacturers get serious about it} and provide the right
mechanisms for securely managing shared processor state. This will
require additions to the ISA that allow any shared state to be either
partitioned or flushed.

\section*{Acknowledgements}
  We would like to thank Dr Stephen Checkoway who helped uncovering documentation 
  on processor functionality.

\label{TextEnd}
\edef\TextEnd{\getpagerefnumber{TextEnd}}
\typeout{TextPages: \TextEnd\space (max \TextPageLimit)}
\ifnum\TextEnd>\TextPageLimit \par\noindent\CComment{red}{{\huge The text is too long---max \TextPageLimit\ pages}}\fi
\balance
{\sloppy
	\footnotesize
    \bibliographystyle{plainnat}
  \bibliography{references}
}
\label{PaperEnd}
\edef\PaperEnd{\getpagerefnumber{PaperEnd}}
\typeout{TotalPages: \PaperEnd\space (max \PageLimit)}
\ifnum\PaperEnd>\PageLimit \par\noindent\CComment{red}{{\huge The paper is too long---max \PageLimit\ pages}}\fi
\end{document}